\documentclass{article}
\usepackage[utf8]{inputenc}
\usepackage[margin=2.7cm]{geometry}
\usepackage{amsmath,amsfonts,amssymb}
\usepackage{graphicx}
\usepackage{multirow,makecell}
\usepackage{adjustbox}
\usepackage{float}
\usepackage[dvipsnames]{xcolor}
\usepackage{ulem}
\usepackage{xurl}
\usepackage{hyperref}
\usepackage{booktabs}
\usepackage{authblk}
\usepackage[square,numbers]{natbib}
\usepackage{caption}
\usepackage[font=footnotesize,labelfont=bf]{caption}

\newcommand{\vi}{``}
\definecolor{Y02A}{HTML}{e68a00}
\definecolor{Y02B}{HTML}{e60000}
\definecolor{Y02C}{HTML}{ff00bf}
\definecolor{Y02D}{HTML}{ff0000}
\definecolor{Y02E}{HTML}{4d004d}
\definecolor{Y02P}{HTML}{0066ff}
\definecolor{Y02T}{HTML}{000099}
\definecolor{Y02W}{HTML}{000000}

\definecolor{Animals}{HTML}{ff5050}
\definecolor{Vegetable}{HTML}{333300}
\definecolor{Fats}{HTML}{e6e600}
\definecolor{Foodstuffs}{HTML}{ff9900}
\definecolor{Mineral}{HTML}{000066}
\definecolor{Chemicals}{HTML}{800000}
\definecolor{Plastics}{HTML}{6699ff}
\definecolor{Leather}{HTML}{cc0066}
\definecolor{Wood}{HTML}{663300}
\definecolor{Paper}{HTML}{cc33ff}
\definecolor{Textiles}{HTML}{3333ff}
\definecolor{Footwear}{HTML}{ff9933}
\definecolor{Stone}{HTML}{99ccff}
\definecolor{Preciousmetals}{HTML}{009999}
\definecolor{Metals}{HTML}{ff0000}
\definecolor{Machinery}{HTML}{660033}
\definecolor{Transportation}{HTML}{ff6600}
\definecolor{Optical}{HTML}{3366ff}
\definecolor{Arms}{HTML}{666699}
\definecolor{Miscellaneous}{HTML}{ff6666}
\definecolor{Art}{HTML}{993333}

\title{The trickle down from environmental innovation to productive complexity}

\author[1]{Francesco de Cunzo}
\author[2]{Alberto Petri}
\author[2]{Andrea Zaccaria}
\author[3,*]{Angelica Sbardella}
\affil[1]{University of Siena, Department of Economics and Statistics, Siena, Italy}
\affil[2]{Institute for Complex Systems, UOS Sapienza, Rome, Italy}
\affil[3]{Enrico Fermi Research Center, Rome, Italy}

\affil[*]{ang.sbardella@gmail.com}

\date{}

\begin{document}

\maketitle
\begin{abstract}
\noindent We study the empirical relationship between green technologies and industrial production at very fine-grained levels by employing Economic Complexity techniques. Firstly, we use patent data on green technology domains as a proxy for competitive green innovation and data on exported products as a proxy for competitive industrial production. Secondly, with the aim of observing how green technological development trickles down into industrial production, we build a bipartite directed network linking single green technologies at time $t_1$ to single products at time $t_2 \ge t_1$ on the basis of their time-lagged co-occurrences in the technological and industrial specialization profiles of countries. Thirdly we filter the links in the network by employing a maximum entropy null-model. In particular, we find that the industrial sectors most connected to green technologies are related to the processing of raw materials, which we know to be crucial for the development of clean energy innovations. Furthermore, by looking at the evolution of the network over time, we observe that more complex green technological know-how requires more time to be transmitted to industrial production, and is also linked to more complex products.

\end{abstract}
\section{Introduction}

The impact of human systems of production and consumption on the environment is increasingly at the center of public debate\cite{ec2019towards,wef2018global,mcmichael2006climate}. As countries face the transition to a more sustainable economy, they will need to take advantage of the new business opportunities and identify profitable entry points in which they can compete in emerging green markets. In particular, there is a broad consensus that green technology development will play a crucial role in sustaining this process as well as in addressing climate change\cite{Popp2010,SternReview}. These are complex and multi-faceted phenomena. The reductionist view of general equilibrium economics will not be able to disentangle the underlying mechanisms and configurations\cite{foxon2012towards}. We thus argue that a framework rooted in the Economic Complexity (EC) literature is better suited to account for the dynamic nature of the socio-economic transformations and structural change that the sustainability transition will ensue, as some exploratory but promising attempts have proven\cite{barbieri2021regional, mealy2020economic,napolitano2020green,sbardella2018green}.\\
In the present paper we propose a novel application of the EC toolbox that allows us to provide a multi-level analysis on the trickle down from single green technological innovations, as proxied by patenting activity in climate change adaptation and mitigation technologies (CCMTs), to industrial production at the level of single products, as proxied by export data \cite{saltarelli2020export}. In order to do so, we draw from studies on the coherence in firm-level patenting\cite{breschi2003knowledge, boschma2013emergence}, on the product space and multi-layer networks\cite{hidalgo2007product,zaccaria2014taxonomy,pugliese2019unfolding}, and also from the work of Sbardella et al.\cite{sbardella2018green}, Napolitano et al.\cite{napolitano2020green} and Barbieri et al.\cite{barbieri2021regional}, who proposed measures of Economic Fitness and technology space based on green technologies. More in detail, we build a network linking single CCMTs, identified through the Y02 Cooperative Patent Classification (CPC) technology class, to single exported products by contracting over the geographical dimension two bipartite networks connecting countries with green technologies at time $t_1$ and countries with exported products at time $t_2 \ge t_1$ respectively, with a time lag between these two layers of $\Delta T \equiv t_2-t_1$ (that could also be zero). This firstly enables us to identify the co-occurrences in the same country of competitive patenting and export, secondly to assess the statistical significance of the co-occurrences via an \textit{ad hoc} maximum entropy null-model\cite{saracco2017inferring}, and finally to define a green technology-product bipartite network, where each link represents the (statistically significant) probability that being proficient in a green technology $\tau$ at time $t_1$ will lead to the successful export of product $\pi$ at time $t_2$. An important feature of the network is its time-dependency: the direction and magnitude of the information flow can change in time, especially when considering different lags between the two original bipartite networks. Each link from a green technology to an exported product highlights the fact that they share similar underlying technological and productive capabilities, therefore indicating the existence of high probability of jumping from the green technology to the linked product. Focusing on the complementary and interrelation between green technological development and specific production lines allows us to identify the green footprint of each product and target specific areas of potential in the green race, providing a valuable external validation for the connection of single products to environmentally-relevant processes otherwise not detectable.\\
As mentioned above, the methodology we propose draws from the EC literature and in particular is based on the Economic Fitness and Complexity (EFC) approach\cite{Tacchella2012,cristelli2013measuring,zaccaria2014taxonomy}. EFC is part of the burgeoning literature on EC\cite{hausmann2007you,hidalgo2009building} and is a multidisciplinary approach to economic big data where the informational content of different types of empirical networks is maximized by using \textit{ad hoc} algorithms which optimize the signal-to-noise ratio.  It has proved highly successful in forecasting\cite{tacchella2018dynamical} as well as explaining\cite{sbardella2018role} economic growth, and was recently adopted by the World Bank\cite{WorldBankEFC} and the European Commission\cite{EurComEFC}. By combining insights from the evolutionary\cite{nelson1982evolutionary, dosi1994introduction} and structuralist approaches\cite{hirschman1958strategy,prebisch1950economic} in economics, EC describes the economy as an evolutionary process of globally interconnected ecosystems and, in a departure from standard economic views, goes beyond aggregate indicators and measures of productive inputs. It considers instead a more granular and structural view of the productive capabilities of an economy by emphasizing the importance of specialization patterns for long-run growth\cite{cristelli2017predictability,hausmann2006growth,hausmann2007you}. One of the most successful fields of application of the EC framework has been the study of local or national innovation systems. Looking through the lenses of EC at the geographical distribution, quality and relatedness of the innovative activities in which economic actors specialize into, as proxied by patent data, allows one to characterize firms\cite{breschi2003knowledge, nesta2005coherence,pugliese2019coherent}, regions or cities\cite{boschma2013emergence,balland2017geography,kogler2013mapping,sbardella2021behind}, as well as to uncover emergent technology patterns at different scales of analysis\cite{napolitano2018technology}. Recently, some promising attempts to draw insights from the EC literature to analyse environmental issues have been put forth, with a special focus on environmental products\cite{fankhauser2013will,hamwey2013mapping,mealy2020economic} and technologies\cite{barbieri2021regional,napolitano2020green,perruchas2019specialisation,sbardella2018green}, setting the basis for a study of the productive or technological capabilities that are relevant to the green economy.\\
Bearing in mind the benefits and the shortcomings of using patent data for studying technological innovation  \cite{arts2013inventions, griliches1998patent,lanjouw1998count}, the choice to study green patenting is motivated by the fact that there is a broad consensus among academics and policy makers that accelerating the development of far-reaching green technologies and promoting their global application are crucial steps, albeit not the only ones, towards containing and preventing greenhouse gas (GHG) emissions and implementing the sustainability transition\cite{oecd2011summary,Popp2010,SternReview}. Despite being a relatively recent phenomenon still at early stages of the life-cycle\cite{barbieri2020knowledge,oecd2011}, over recent years we have witnessed a great acceleration in the development of green technologies, especially in the energy and transport area\cite{oecd2011summary}. These technologies show distinctive features with respect to non-green ones. They appear to be heterogeneous, encompassing many domains of know-how\cite{Barbieri2020b} across different geographical areas\cite{sbardella2018green}, but are linked in non-trivial ways to the pre-existing knowledge base\cite{barbieri2020knowledge,barbieri2021regional}.
However, it is important not to disregard the intrinsic limits and difficulties of a \vi big technological fix" \cite{parkinson2010coming,sarewitz2008three} and to be aware that science and technology can indeed provide effective tools to tackle the climate crisis, but they will be the more effective the more they will be accompanied by a project of radical transformation of current production and development models\cite{ec2018communication,ec2019towards}.\\
Within this context, our findings allow several considerations to be made. First, and somewhat surprisingly, the products with the highest green technology footprint -- i.e., most connected to green technologies in the network-- concern the export of raw material products, such as mineral, metal and chemical products. Their persistent presence and importance in our network resonate with the literature on the raw material requirements that the green transition will entail\cite{EuCRM2020, Worldbank2020, IEA2021, Valero2018}. In fact, materials like lithium, cobalt, indium, nickel and many others are key inputs for several green technologies, in particular for those related to renewable energy and electrical mobility. Hence, to deal with the climate and environmental crisis, the extent to which an increase in the development of green technologies could affect mineral demand will need to be carefully taken into consideration when countries or international organisations\cite{EuGreenDeal, ec2019towards} take action or reflect upon future scenarios. Among the goods that according to our analysis are significantly related to green technologies we also find different products related to the export of animal and vegetable products -- which are mostly connected  to technologies for GHG capture and storage -- and machinery and electrical products -- which especially show connections with CCMTs in information and communication technologies. 
Finally, another key result of our analysis is that the links in the green technology-exported product network structure change when we increase the time lag between between green patenting and product exports. When passing from the simultaneous observation of the two network layer, to a 10 year time lag between the two, we observe an increasing number of links between complex green technologies and complex products, suggesting that more complex green technological know-how requires longer to unfold into industrial production and to enter in connection with more sophisticated production lines.

\section{Results}\phantomsection\label{sec:result}

As mentioned above, the aim of this paper is to leverage statistically validated networks to explore the connections between green technologies and exported products, i.e. the trickle down from green technology innovation to industrial production. Each link between a green technology and a product not only indicates that they require similar capabilities to be competitive in them, but also that having a comparative advantage in a green technology is a good predictor for the development and export of a specific product. We compute the validated links for two different aggregations of the data on exported products, moving from a broader level of description --- consisting of 97 so-called product chapters, labeled with 2-digit codes --- to a more detailed one --- consisting of 5053 product subheadings, labeled with 6-digit codes. Moreover, we are able to assess the evolution of the green technology-product network by taking into account the effect of a time lag of 10 years between the development of green technologies and the export of products. 

\phantomsection
\subsection{Green techs - products connections: general remarks}
\phantomsection
\subsubsection{Aggregated analysis}

\begin{figure}[hptb]
\centering
\includegraphics[width=\linewidth]{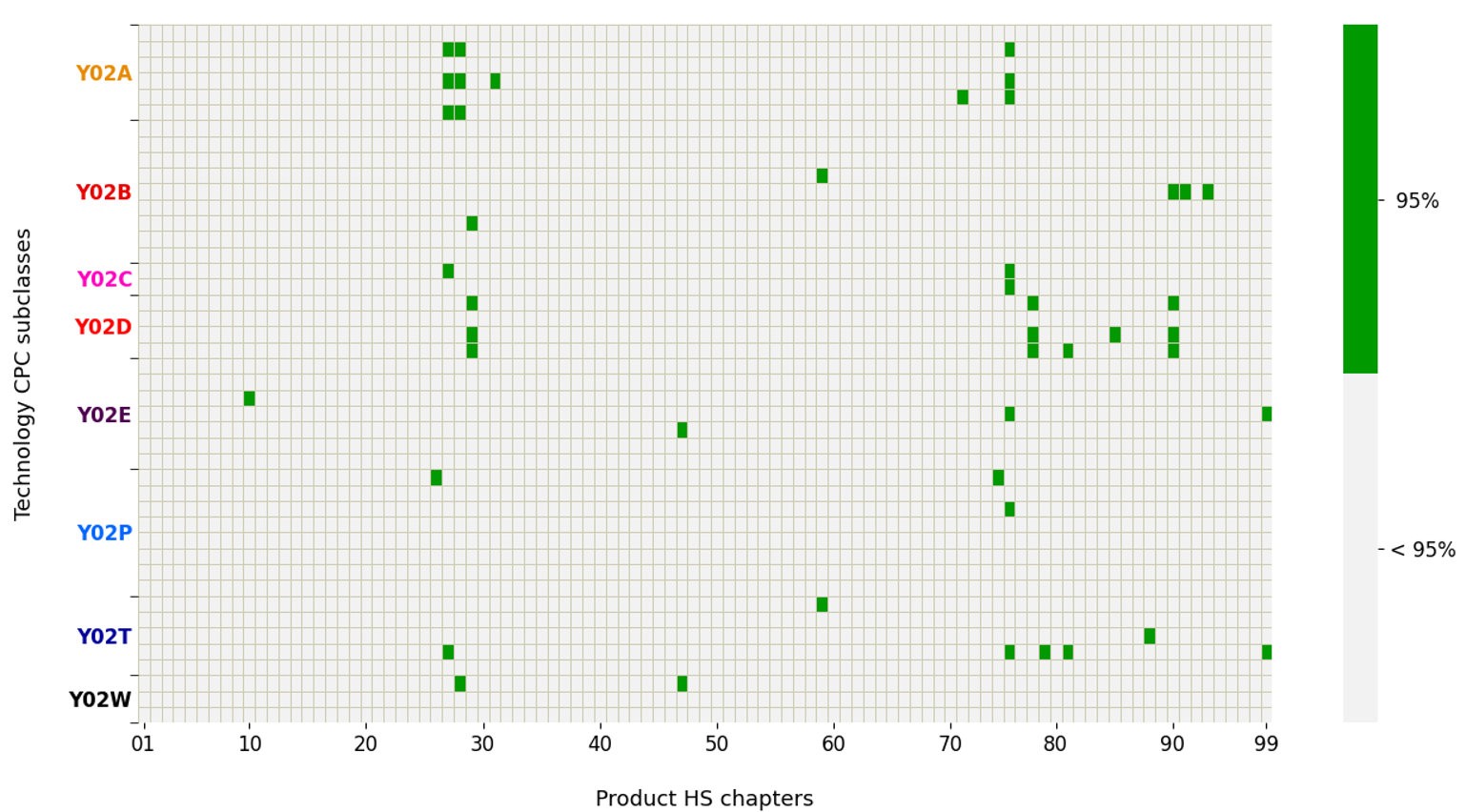}
    \caption{Heatmap representation of network links at 95\% level of significance. Y-axis = CPC codes of green technology sub-classes; X-axis = 2-digit exported products. Each green rectangle corresponds to a link between the corresponding green technology on the y-axis and exported product on the x-axis.}
    \label{Fig1}
\end{figure}

\begin{figure}[hptb]
\centering
\includegraphics[width=\linewidth]{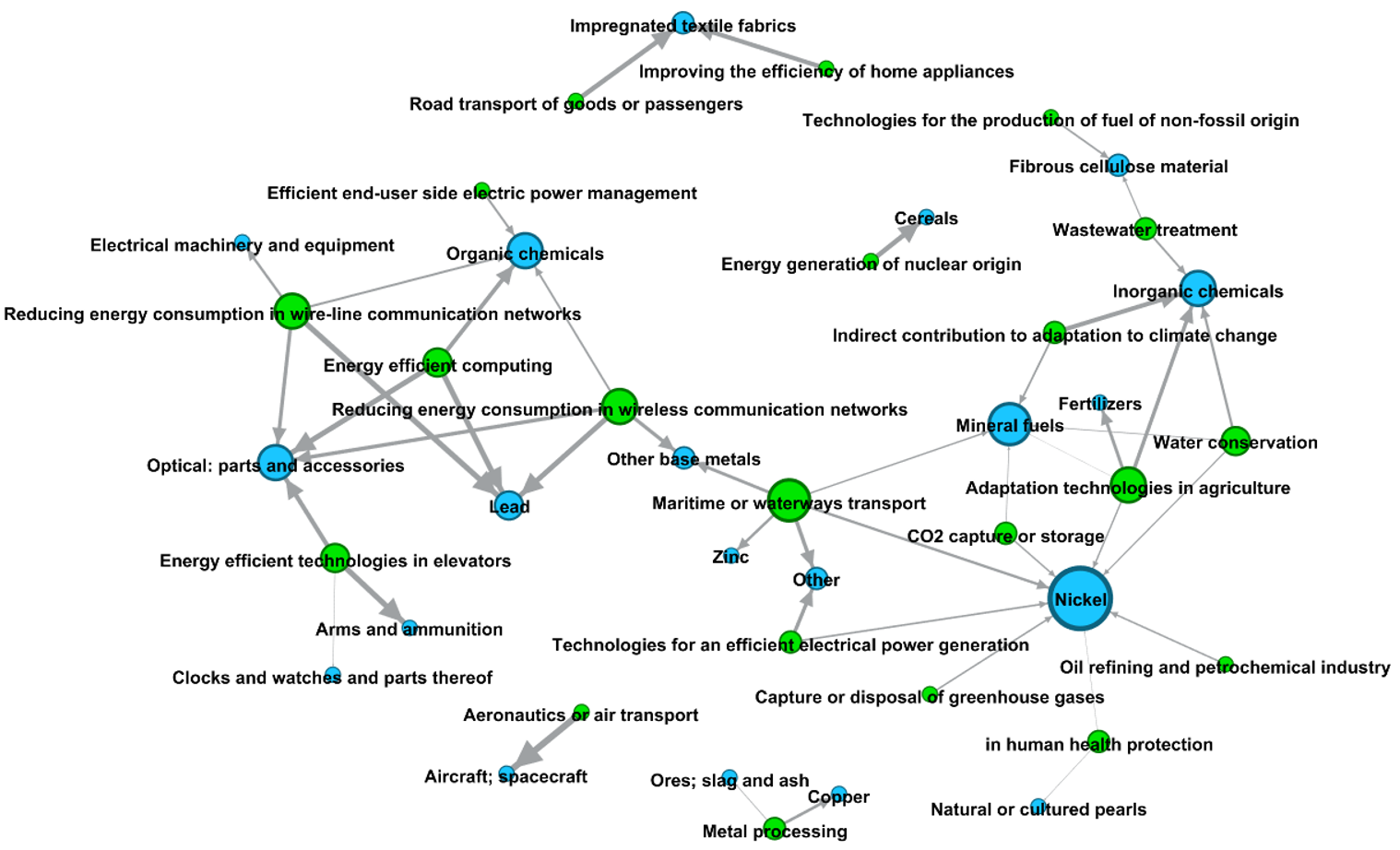}
    \caption{Directed network from green technologies to exported products for a time lag $\Delta T = 0$. Products aggregation: 2-digit level. Nodes' size depends on their degree;  edges are weighted according to the value of the Assist matrix $A_{\tau\pi}$.}
    \label{Fig2}
\end{figure}

In order to build the multi-layer network in which green technologies are linked to exported products, we start by considering two binary networks: the first one connects countries to the green technologies they patent competitively, the second one connects countries to the products they export competitively. By summing over the geographical dimension we then build the so-called \textit{Assist Matrix}\cite{pugliese2019unfolding, zaccaria2014taxonomy}, which is the adjacency matrix of the multi-layer network connecting green technologies to exported products, in the following way:
\begin{equation}
\label{assist1}
A_{\tau,\pi}(t_{1},t_{2})= \dfrac{1}{u_\tau(t_{1})} \sum_c\dfrac{M_{c\tau}(t_{1}) M_{c\pi}(t_{2})}{d_c(t_{2})}, \text{ with } \begin{cases}
    d_{c}(t_{2}) = \sum_{\pi'}M_{c\pi'}(t_{2})\\\\
    u_{\tau}(t_{1})= \sum_{c'}M_{c'\tau}(t_{1})
\end{cases}
\end{equation} 
where the $\textbf{M}$ matrices define the bipartite networks where countries are linked to the green technologies or exported products in which they have a comparative advantage (see \nameref{sec:methods}). That is, we are counting suitably normalized co-occurrences, the normalization factors being the product diversification of country $c$ at year $t_{2}$ $d_{c}(t_{2})$ -- i.e. the number of products included in the export basket of that specific country -- and the ubiquity of the green technology $\tau$ at year $t_{1}$ $u_{\tau}(t_{1})$ -- i.e. the number of countries that are patenting in that specific technological sector. The resulting green technology-product links are then statistically validated by using the Bipartite Configuration Model\cite{squartini2011analytical,saracco2017inferring}. We details of the procedure can be found in the \nameref{sec:methods} section.\\
We start our analysis by considering simultaneous normalized co-occurrences, that is with a time lag $\Delta T \equiv t_{2} - t_{1} = 0$ between the two network layers. Firstly, we investigate the links between green technologies and exported products at a 2-digit aggregation level. Figure \ref{Fig1} represents the adjacency matrix of the green technology-product network at a 95\% statistical significance, where we find  46 significant links in total (i.e. 46 green rectangles in the figure). This figure allows us to provide some initial qualitative insights on which green technologies and exported products are connected and which are not. As regards green technologies we note that, although not uniformly, all technology sub-classes (see Table \ref{Tab1} for CPC Y02 code descriptions) have some links to products and are present in the network. The same cannot be said for the exported product layer: some 2-digit product sections are almost completely disconnected, including e.g. \textit{Foodstuffs, Plastics/Rubbers, Leather} and \textit{Textiles}, while others have a considerable amount of links. In particular, product like \textit{Mineral fuels, Nickel, Lead, Organic} and \textit{Inorganic chemicals} are highly connected with green technologies such as   \textit{technologies for adaptation to climate change} (Y02A) and \textit{CCMTs in information and communication technologies} (Y02D), indicating that a relatively high number of countries are active in both. This hints at an overlapping of the green technological know-how and the productive capabilities needed for being proficient in both, suggesting that countries that do patent in technology sub-classes as Y02A and Y02D not only are more likely to export raw material products, but also that different types of metals and chemicals are highly connected to R\&D in CCMTs, and thus new sustainable avenues in their production could be explored. The topic of raw material products and a specific case study will be discussed in more detail below.\\
In Fig. \ref{Fig2} we offer an alternative representation in which we show the directed network between green technologies and exported products, with the node size being proportional to the node degree. The size of the edges also varies between links: each edge linking two nodes is more or less thick according to the corresponding Assist Matrix entry. The network representation permits a clear distinction between the disconnected components (such as the two nodes relative to air transport at the bottom left) and the large connected component in the center. For instance, it is interesting to notice the energy-related cluster on the left portion of the plot, where green technologies aimed at improving efficiency in computing, in wire-line and wireless communication networks and in the electric power management are linked to the export of raw material products and optical and electrical products, which are important inputs for this kind of technologies.

\begin{table}[hptb]
\centering
\resizebox{\textwidth}{!}{%
\begin{tabular}{@{}ll@{}}
\toprule
\multicolumn{1}{l}{\textbf{Class or Sub-class}} & \multicolumn{1}{l}{\textbf{Title and description}} \\ \midrule
\textbf{Y02} & \begin{tabular}[c]{@{}l@{}} \textbf{TECHNOLOGIES  OR APPLICATIONS FOR MITIGATION OR ADAPTATION} \\ \textbf{AGAINST CLIMATE CHANGE}\end{tabular} \\ \midrule
\textcolor{Y02A}{Y02A} & Technologies   for adaptation to climate change \\ \\
\textcolor{Y02B}{Y02B} & \begin{tabular}[c]{@{}l@{}}Climate change   mitigation technologies related to buildings, e.g. housing, house appliances or \\ related end-user applications, including the residential sector\end{tabular} \\ \\
\textcolor{Y02C}{Y02C} & Capture, storage,   sequestration or disposal of greenhouse gases \\ \\
\textcolor{Y02D}{Y02D} & \begin{tabular}[c]{@{}l@{}}Climate change   mitigation technologies in information and communication technologies,\\  i.e. information and communication technologies aiming at the reduction of their own energy use\end{tabular} \\ \\
\textcolor{Y02E}{Y02E} & \begin{tabular}[c]{@{}l@{}}Reduction of   greenhouse gas (GHG) emissions, related to energy generation, \\ transmission or   distribution, including renewable energy, efficient combustion, \\ biofuels, efficient transmission and distribution, energy storage, and hydrogen technology\end{tabular} \\ \\
\textcolor{Y02P}{Y02P} & Climate change   mitigation technologies in the production or processing of goods \\ \\
\textcolor{Y02T}{Y02T} & Climate change   mitigation technologies related to transportation, e.g. hybrid vehicles \\ \\
\textcolor{Y02W}{Y02W} & Climate change   mitigation technologies related to wastewater treatment or waste management \\ 
  \bottomrule
\end{tabular}%
}
\caption{CPC Y02 tagging scheme. Source: EPO\cite{epoY}. In the first column the CPC code identifying the Y02 technology sub-class is reported. The second column reports the corresponding description.}
\label{Tab1}
\end{table}

\subsubsection{Fine-grained connections}
\begin{figure}[hptb]
\centering
\includegraphics[width=.95\linewidth]{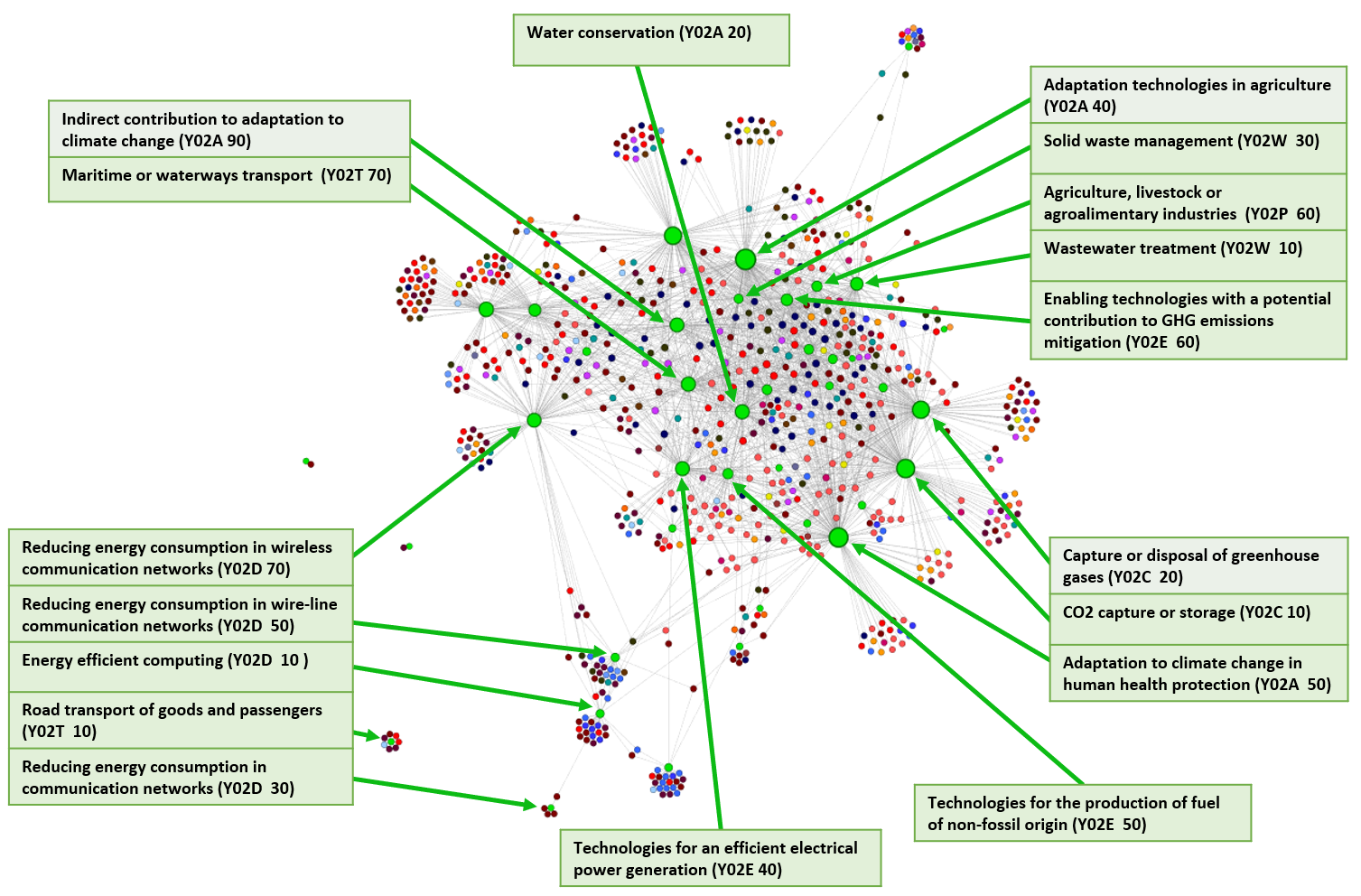}
    \caption{Directed network from green technologies to exported products for a time lag $\Delta T = 0$. Products aggregation: 6-digit level. Nodes' size is proportional to their degree. Green nodes: green technologies. Green arrows link the description to some of them. All the other nodes: exported products (coloured according to Table \ref{Tab2}).}
    \label{Fig3}
\end{figure}

\indent We move forward into the analysis by considering the 5053 exported products present in the classification at 6-digit aggregation level. Increasing the level of data breakdown reveals the potential of our methodology, that can be easily applied to any level of data aggregation, and when applied to fine grained information can provide very punctual insights. Figure \ref{Fig3} represents the entire bipartite green technology-product network. The dimension of the nodes is proportional to their degree; the green ones correspond to the green technologies, while all the others correspond to the exported products, and are coloured according to the product sections they belong to (see Table \ref{Tab2}). We notice that, in line with the 2-digit product case, almost all green technologies are present: indeed, 39 (out of the total 44) are present in the network. This means that almost all green technologies are connected to the production of at least one product. However, depending on where the green nodes are placed in the network, a green technology may be more or less integrated into the production system as a whole. More specifically, we can see that the periphery of the network is dominated by technologies related to services and transport, while the core of the network contains technologies belonging to sub-classes like Y02A, which covers technologies for the adaption to the adverse effects of climate change in human, industrial (including agriculture and livestock) and economic activities, and Y02W, which covers CCMTs related to waste management.\\
In Table \ref{Tab2} we collect some descriptive information on the distribution of product nodes and edges in the network. More in detail, products belonging to primary sectors such as animal and vegetable products show many connections with green technologies. In particular, the links we observe are between green technologies and the export of meat, fish, milling industry products and miscellaneous grains. All of these are largely connected with Y02A --- especially with Y02A 40 - \textit{adaptation technologies in agriculture, forestry, livestock or agroalimentary production} and Y02A 50 - \textit{in human health protection} --- and with \textit{technologies for capture, storage, sequestration or disposal of GHG} --- i.e. Y02C. This is consistent with the high level of pollution and emissions that the agricultural sector is accountable for\cite{owidghgemissions}. Finally, in line with the results obtained in the 2-digit product case, the subheadings belonging to  minerals, chemicals and metals product sections are confirmed to be highly connected to green technologies. We elaborate on this by focusing on the export of a specific product in the following.

\begin{table}[hptb]\centering
\resizebox{.95\textwidth}{!}{%
\begin{tabular}{|l|l|l|l|l|}
\hline
\textbf{Product Section}                                            & \textbf{\begin{tabular}[c]{@{}l@{}}2-digit\\ included\end{tabular}} & \textbf{\begin{tabular}[c]{@{}l@{}}\# of 6-digit \\ products (\%)\end{tabular}} & \textbf{\begin{tabular}[c]{@{}l@{}}\# of nodes in \\ the network (\%)\end{tabular}} & \textbf{\begin{tabular}[c]{@{}l@{}}\# of edges in \\ the network (\%)\end{tabular}} \\ \hline
{\color{Animals} \textbf{Animal \& animal products}} & 01-05    & 228 (4.5\%)  & 114 (15.6\%)  & 424 (19.6\%) \\ \hline
{\color{Vegetable} \textbf{Vegetable products}}& 06-14  & 256 (5.1\%)  & 54 (7.4\%)  & 151 (7.0\%)   \\ \hline
{\color{Fats} \textbf{Fats, oils and waxes}} & 15 & 45 (0.9\%) & 12 (1.6\%)   & 35 (1.6\%)    \\ \hline
{\color{Foodstuffs} \textbf{Foodstuffs}}& 16-24& 193 (3.8\%)& 30 (4.1\%)& 61 (2.8\%)    \\ \hline
{\color{Mineral} \textbf{Mineral products}}& 25-27& 148 (2.9\%)& 58 (8.0\%)& 355 (16.4\%)  \\ \hline
{\color{Chemicals} \textbf{Chemicals \& allied industries}} &28-38& 789 (15.6\%)& 124 (17.0\%)& 295 (13.6\%)  \\ \hline
{\color{Plastics} \textbf{Plastics/Rubbers}}& 39-40& 211 (4.2\%)& 8 (1.1\%)& 11 (0.5\%)    \\ \hline
{\color{Leather} \textbf{Leather}}& 41-43& 69 (1.4\%)& 14 (1.9\%)& 38 (1.8\%) \\ \hline
{\color{Wood} \textbf{Wood}}& 44-46& 93 (1.8\%)& 16 (2.2\%)& 42 (1.9\%)    \\ \hline
{\color{Paper} \textbf{Paper}}& 47-49& 144 (2.9\%)& 30 (4.1\%)& 103 (4.8\%)   \\ \hline
{\color{Textiles} \textbf{Textiles}}& 50-63& 801 (15.9\%)& 24 (3.3\%)& 42 (1.9\%)    \\ \hline
{\color{Footwear} \textbf{Footwear/Headgear}}& 64-67& 49 (1.0\%)& 2 (0.3\%)& 2 (0.1\%)     \\ \hline
{\color{Stone} \textbf{Stone/Glass}}& 68-70& 143 (2.8\%)& 11 (1.5\%)& 17 (0.8\%)    \\ \hline
{\color{Preciousmetals} \textbf{Precious stones and metals}}& 71& 53 (1.1\%)& 24 (3.3\%)& 69 (3.2\%)    \\ \hline
{\color{Metals} \textbf{Metals}}          & 72-83& 568 (11.2\%)& 94 (12.9\%)& 326 (15.1\%)  \\ \hline
{\color{Machinery} \textbf{Machinery/Electrical}}& 84-85& 769 (15.2\%)& 58 (8.0\%)& 91 (4.2\%)    \\ \hline
{\color{Transportation} \textbf{Transportation}}& 86-89& 131 (2.6\%)& 17 (2.3\%)& 25 (1.1\%)    \\ \hline
{\color{Optical} \textbf{Optical instruments}}& 90-92& 217 (4.3\%)& 31 (4.3\%)& 59 (2.7\%)    \\ \hline
{\color{Arms} \textbf{Arms and ammunition}}& 93& 20 (0.4\%)& 4 (0.6\%)& 10 (0.5\%) \\ \hline
{\color{Miscellaneous} \textbf{Miscellaneous manufactured articles}} & 94-96& 118 (2.3\%)& 1 (0.1\%)& 1 (0.1\%)     \\ \hline
{\color{Art} \textbf{Works of art}}& 97& 7 (0.1\%)& 3 (0.4\%)& 6 (0.3\%)     \\ \hline \hline
\textbf{TOTAL}& /& 5052& 729& 2163\\ \hline
\end{tabular}}
\caption{Exported product sections. $1^{st}$ column: product section names; $2^{nd}-3^{rd}$ columns: which 2-digit products and how many 6-digit products are included. $4^{th}-5^{th}$ columns: number of nodes and edges in the network of Fig. \ref{Fig3}. The percentages between parenthesis are computed with respect to the TOTAL values reported in the ending line. Note that product \textit{999999: Commodities not specified according to kind} is not included. }
	\label{Tab2}
\end{table}

\subsubsection{A case study: cobalt}
\begin{figure}[hptb]
\centering
\includegraphics[width = .9\textwidth]{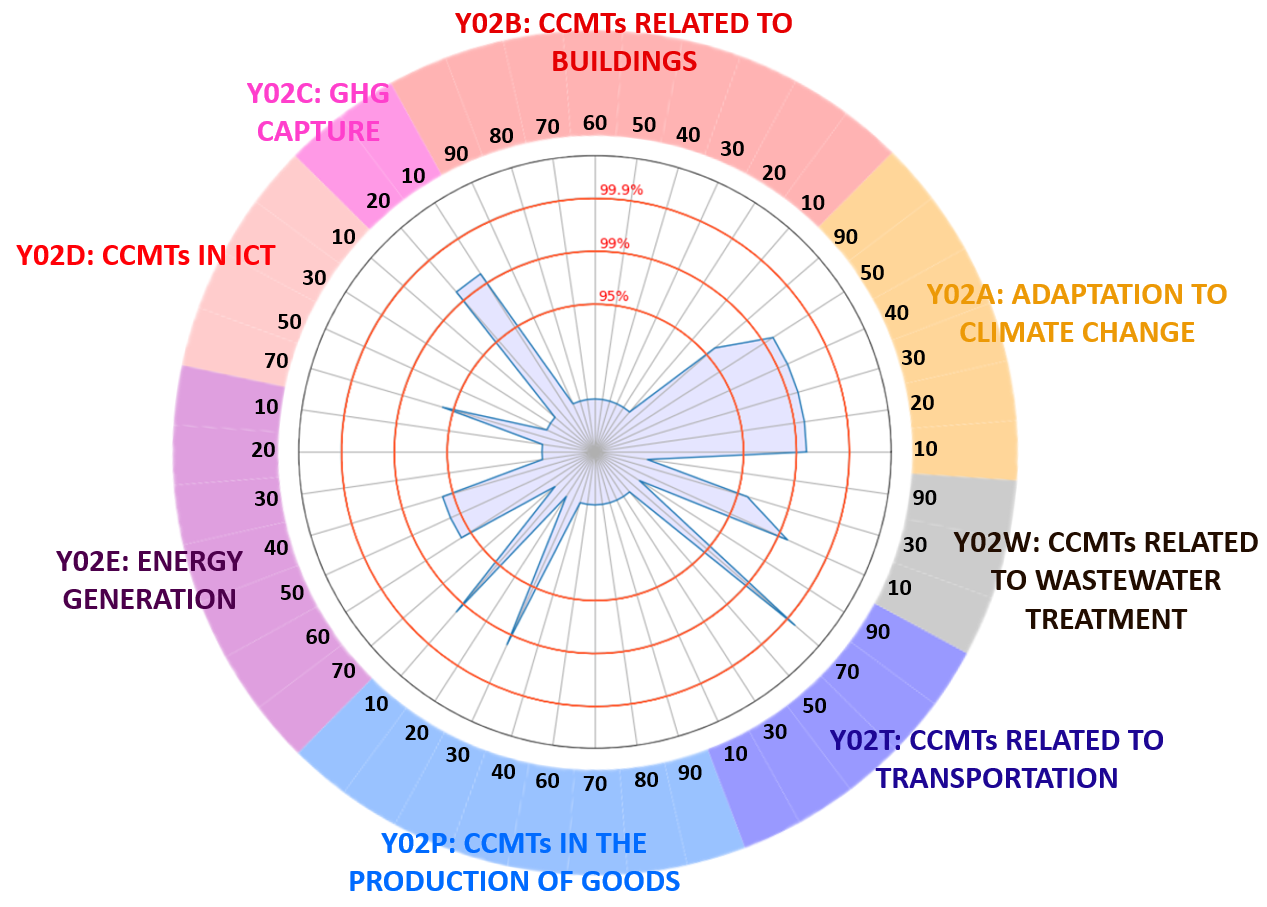}
    \caption{Focus on the export of \textit{cobalt and other intermediate products of cobalt metallurgy} (Harmonized System code 810520). Along the circular border of the figure, the CPC codes of the 44 green technology groups are labelled. Within the figure, three concentric circles delimit the significance levels of 99.9\%, 99\% and 95\%. Each peak in blue that exceeds the level delimited by one of the inner circles corresponds to a link that the cobalt has with the green technology labeled on the border.}
    \label{Fig4}
\end{figure}

An interesting product export example in our green technology-product network is that of cobalt and other intermediate products of cobalt metallurgy. Figure \ref{Fig4} layout highlights which technologies are significant requirements for the successful export of cobalt, with a level of confidence  larger than 95\%. In the figure, three red concentric circles delimit the 99.9\%, 99\% and 95\% level of significance. The blue peaks exceeding one of these circle in the figure denote that the export of cobalt is linked at the corresponding level of significance with the green technology labeled around the circular border. In particular, cobalt export is linked with \textit{technologies for adaptation to climate change} (Y02A), \textit{related to transportation} (Y02T) and \textit{waste treatment} (Y02W), \textit{for energy generation, transmission and distribution} (Y02E), and with \textit{CCMTs in in information and communication technologies} (Y02D) and \textit{in the production or processing of goods} (Y02P).\\
The cobalt example further reinforces what we clearly observe in all the results of the analysis, namely a consistent presence of raw materials among the exported products that are most linked to green technologies. This is not surprising: in fact, an emerging literature on the topic is trying to estimate the mineral intensity of green technologies and to forecast how their spread will shape the mineral demand in the years to come\cite{Worldbank2020, IEA2021, Herrington2021, Valero2018}. In particular, cobalt is considered a high-impact mineral for the clean energy transition. Indeed, to meet expected future demand its production needs to increase up to nearly 500\% of 2018 levels by 2050\cite{Worldbank2020}. Cobalt is a key element in energy storage technologies, which are crucial to a low-carbon transition for two main reasons: they are used in the transport sector to power electric vehicles and they are needed to store energy from other intermittent renewable sources, such as solar photovoltaic and wind. Given that 64\% of global cobalt supply comes from the Democratic Republic of Congo\cite{ECReportBattery}, the risks associated with meeting its demand --- which will rise if certain climate targets are to be met --- and the cross-cutting way in which it is used in green technologies, have led to cobalt being placed on the European Commission's list of critical raw materials\cite{EuCRM2020}, which includes materials considered critical for their supply risk and economic importance. The list is updated every three years, and cobalt has been in it since its first version published in 2011\cite{EuCRM2011}. It is worth to notice that we do not have data on green patents for the Democratic Republic of Congo. However, even if its main world supplier is missing, we still observe many connections between cobalt and cobalt metallurgy products and green technologies. In particular, these connections arise from the co-occurrences of several green technologies and cobalt products export in countries like Australia, Belgium, Canada, Finland, Norway, Russia and South Africa, which are all important producers of cobalt and refined cobalt\cite{BritGeoSurvey2022}.

\subsection{Connections in a 10 year horizon}

\begin{figure}[hptb]
\centering
\includegraphics[width=.75\linewidth]{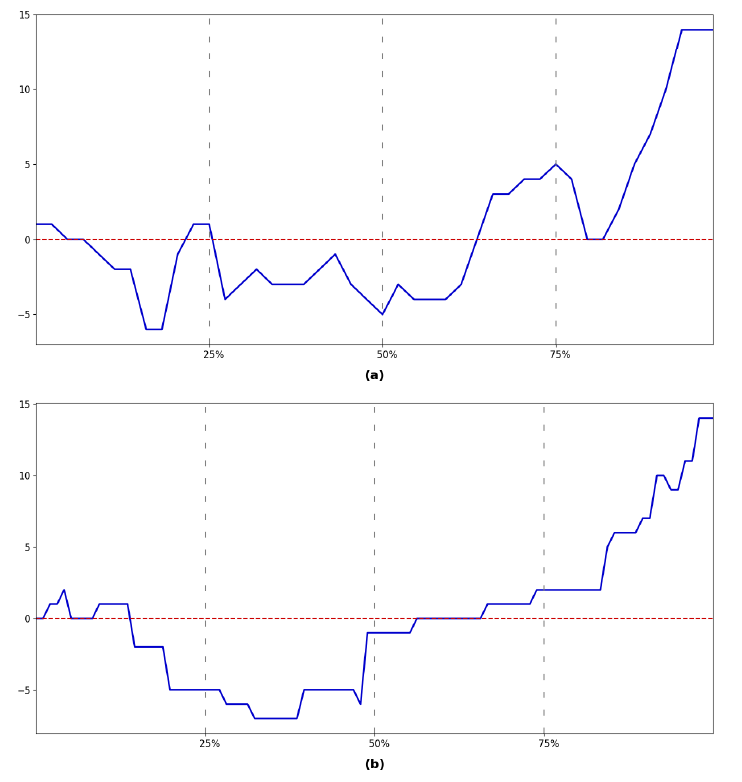}
    \caption{Cumulative difference between the number of node links for the time lag $\Delta T = 10$ and $\Delta T = 0$. Panel (a) refers to green technologies, while panel (b) refers to the 2-digit exported products. In their respective panel, green technologies and exported products are ordered (not labeled) in ascending order of complexity ranking. The labels of "25\%, 50\% and 75\%" delimit the first, second and third quartiles of the complexity ranking (moving from the last position to the first one). If the y-value is below 0 (dashed red line), then the cumulative number of links delimited by the corresponding green technology or product in the x-axis is higher for $\Delta T = 0$. On the contrary, if the y-value is above 0, then the cumulative number of links is higher for $\Delta T = 10$. }
      \label{Fig5}
\end{figure}

With the aim of analysing whether the spectrum of green technologies needed to gain a comparative advantage in a variety of productive sectors changes over time, here we explore how the links between green technologies and exported products change, both in qualitative and quantitative terms, moving from a time lag between the green technology and  exported product layers of $\Delta T \equiv t_{2} - t_{1} = 0$ to $\Delta T = 10$. In fact, our analysis can  be conducted also by considering  different values of $\Delta T$ allowing for a dynamic perspective on the green technology--production nexus.\\
When considering $\Delta T = 10$ from a quantitative point of view we observe a slight increase in the total number of links, both in the case of 2-digit and 6-digit products (from 46 to 60 links in the case of 2-digit products and from 2166 to 2354 links in the 6-digit case). This finding is coherent with the results presented in Pugliese et al.\cite{pugliese2019unfolding}, in which the authors show that technological advancements on average anticipate export. The increase of roughly 10\% of the resulting links suggests that green technologies are better integrated into the production process after a ten years digestion.\\
Regarding possible differences in the properties of the linked technologies and products for both time lags, in Fig. \ref{Fig5} we plot the cumulative increment in the number of links for both green technologies and exported products. In particular, in the x-axis of the two plots we rank green technologies and exported products by increasing complexity, which is computed through the implementation of the Economic Fitness \& Complexity (EFC) algorithm\cite{Tacchella2012} (see the Supplementary Information [Economic Fitness \& Complexity algorithm]). The blue lines in the figures plot the cumulative difference between the number of links that each activity has for $\Delta T = 10$ and $\Delta T = 0$. What emerges from the two plot layouts is particularly significant: the new links that appear when the time lag is increased are relative to more complex products and also more complex green technologies. Therefore, it is likely that more complex potential spillover effects in the economic production deriving from the development of a green technology will arise at a later stage. This is in line with the idea that more complex green technological know-how requires more time to be transmitted to the productive sectors. Moreover, this finding is in agreement with Barbieri et al.\cite{barbieri2020knowledge,Barbieri2020b} that study the relationship between green and non-green knowledge bases and argue that green technologies are generally complex and have a heterogeneous development process, involving different domains of know-how. 

\section{Discussion}\phantomsection\label{sec:discussion}

To address the climate crisis, it is necessary to change the way economies have grown and developed in recent decades, as this has led to an overall ecological overshoot. To overshoot means that humanity is beyond the limits that the planet imposes to our economy: we are using more resources and producing more waste than can be regenerated and absorbed without consequences\cite{Limitstogrowth}. Different approaches can be adopted to steer economies onto a more sustainable path. For example, a strand of research focuses on the analysis and development of green growth policies that aim at promoting economic growth by mitigating its environmental impact through the decoupling of growth and greenhouse gas (GHG) emissions\cite{oecd2011summary}. In contrast, in the literature on degrowth it is argued that such decoupling is not feasible, and that a rethinking of consumption and production patterns is necessary to address the climate crisis\cite{DAlessandro2020}. Therefore, the policies that characterise these approaches are not always complementary; on the contrary,  they often arise in opposition to each other. However, even across different approaches there is agreement on some essential actions that should be undertaken in any case: among these is certainly the development of environmental innovations aimed at reducing GHG emissions. This is where our article comes in. In fact, through our work we are able to establish at a very detailed level which productive activities benefit the most from the development in green innovation. Therefore, our focus is on possible industrial scenarios resulting from the development of green technologies. In particular, we discuss how green technological know-how is transmitted to industrial production at the product level, even years later. However, we are in no way arguing that there is a causal relationship that links green patenting to subsequent product export, we are just observing statistical significant probabilities that having a comparative advantage in a green technology will lead to the export of a specific product.\\
Among our main findings, we emphasised the presence of many links between green technologies and the export of raw materials, especially mineral and metal products. In addition, we provide evidence on the presence of significant connections between products belonging to the agricultural sector, like \textit{Animal \& Animal products}, and green technologies aimed at the capture and storage of GHG emissions. Finally, we observe that as the years between the the filing of a green patent and the export of a product increase, so does the complexity, and therefore the skills and expertise required, of the products and technologies that are linked together. Throughout the paper we have argued about how raw materials are necessary for the development of environmental technologies: in recent years, several reports analysing this issue have been published by international organisations and institutions\cite{Worldbank2020,IEA2021,EuCRM2011}. Therefore, our paper strongly stands in this context: we claim that in order to spread the development of green technologies and to increase their use with the objective of achieving a sustainable transition of the economies, the raw materials intensity of these technologies is a core issue to be deepened. Indeed, it is important to plan appropriate strategies to meet the expected surge in raw materials demand, or to reduce the supply dependency on individual countries that could undermine the stability of the overall raw materials value chain. Despite the fact that these materials are considered as necessary inputs for the realisation of green technologies, thus suggesting an inverse relationship to that studied by us, we nevertheless believe that the links from green technologies to mineral product exports we observe have a strong relevance. Future research could thus explore this issue further, for instance by looking at the connections from exports to green patenting, or by considering import data.\\
We believe that our results are particularly relevant for a number of reasons. First of all, being able to go into such detail in assessing the implications that emerge from the development of green technologies, not only evaluating their collective impact on industrial production, but discussing individual product exports with individual technology domains on a case-by-case basis, has very strong policy implications. For example, such an analysis could provide support for the industrial policies of a given country, even in the long term, by looking at the patent portfolio in which it is currently competitive. In addition, possible contributions could be made to the classification of environmental products: some products could be linked to green technologies because of the low environmental impact of their production processes. Being able to monitor the export and import of environmentally sensitive products is a central objective on the global policy agenda. For example, the Harmonized System under which exported products are classified is about to be updated\cite{steenblik2020code} with the main changes being the introduction of several 6-digit subheadings that include environmental goods in order to facilitate their trade.\\
For future developments in the analysis, we believe it would be important to take in consideration time intervals of more years than those considered in our dataset. The export data update just mentioned could be very useful in this respect, as it would add more annual collections to the product dataset which in turn would allow us to increase the time lag of our analysis even beyond 10 years. We expect that this would lead to an increase in the signal from green technologies to products, as previous analysis shows that the peak of the technological impact on industrial production is reached after about 20 years\cite{pugliese2019unfolding}. Finally, another important aspect could extend the layers of activities, and consequently the type of data, taken into account in the analysis: for example, by including also data on employment and wages at a sector or occupational level, or on the scientific production of countries, we could broaden our understanding of how the production and technological structure of a country or a region can make the transition to new green sectors.

\section{Methods}\phantomsection\label{sec:methods}
\subsection{Data}
We use data on patent applications in environment-related domains as a proxy for environment-related innovation and on exported products as a proxy for economic production \cite{saltarelli2020export}. Both datasets consist of single data collections recorded annually at a country level. In particular, we have information on patent applications on 44 green technological fields --- corresponding to the CPC groups listed in the Supplementary Information [Table S2: CPC detailed descriptions]  --- for 48 countries between 1995 and 2019 and on product exports --- whose number depends on the level of aggregation considered: 97 in the 2-digit case, 5053 in the 6-digit one --- measured in US dollars for 169 countries between 2007 and 2017. As explained in detail in the next section, our methodology requires selecting the countries in common between the two data collections, which turn out to be 47. All data can be represented as matrices: in particular, we denote by $\textbf{W}(t)$ and $\textbf{V}(t)$ the matrices corresponding respectively to the data of green patents and exported products in year $t$. Each matrix has a number of rows and columns equal to the number of countries $c$ and activities $a$ respectively, where the latter refer to both green technologies $\tau$ and exported products $\pi$. We report more detailed description of the two datasets we use, including also a complete list of all countries at our disposal, in the Supplementary Information [Data features].

\subsection{Data preprocessing}
\phantomsection
\subsubsection{Temporal aggregation} \label{Aggregations}

Both the export of products and the patenting activity are collected yearly: it is then possible to investigate the connections on different time scales. While annual data can offer more detailed results, i.e. distinct for each year considered, it may also supply them with more noise. In fact, data can fluctuate from one year to another. In order to minimize the possibility that the green technology-product connections arise from data fluctuations, we consider the total volume of products and patents produced in given time intervals. For our analysis, we compute the matrices $\textbf{W}(\delta,t)$ and $\textbf{V}(\delta,t)$, corresponding to the time interval of $\delta$ years ending in the year $t$. To this aim, we sum the yearly matrices $\textbf{V}(t)$ and $\textbf{W}(t)$ over the year range $\delta$:
\begin{eqnarray}\label{sumover5years}
\begin{aligned}
\textbf{V}(\delta,t) &= \sum_{t'=t-\delta+1}^{t} \textbf{V}(t')\\
\textbf{W}(\delta,t) &= \sum_{t'=t-\delta+1}^{t} \textbf{W}(t')
\end{aligned}
\end{eqnarray}
Summing data over a time window of $\delta$ years reduces the noise in our results, giving more weight to patents and exports that are consistently registered several times in nearby years.\\
Given the years of the datasets at our disposal, we decide to sum the matrices over 5 years ($\delta = 5$). Starting from the layer of exported products, we select the two most recent 5-year aggregate matrices available to us, with the condition that the years included did not overlap each other. Therefore, these matrices are $\textbf{V}(\delta,t)=\{\textbf{V}(5,2012); \textbf{V}(5,2017)\}$. Then, depending on which time lag $\Delta T$ we consider between the two layers, we select the green patents matrices. So, for a time lag $\Delta T = 0$, the corresponding matrices are $\textbf{W}(\delta,t)=\{\textbf{W}(5,2012); \textbf{W}(5,2017)\}$, while for $\Delta T = 10$ they are $\textbf{W}(\delta,t)=\{\textbf{W}(5,2002); \textbf{W}(5,2007)\}$ so that the green technologies "anticipate" the exports. To easy the notation, from now on we do not express the $\delta$ dependency of the data matrices, and all our results are produced from the analysis of the aggregated 5-year data collections just mentioned. We have conducted robustness tests of the links we found with respect to changes for both different aggregation time intervals $\delta$ and the final year $t$. We report this tests in the Supplementary Information [Robustness test]. The green technology-product links we find are robust to such changes in the parameters.

\subsubsection{Revealed Comparative Advantage}

Both exports and patents matrices strongly depend on the total size of the economy or sector. In order to remove this correlation, which hides the capability content of these activities, we computed the revealed comparative advantage (RCA)\cite{balassa1965trade} . The RCA is computed as the ratio between the weight of activity $a$ (be it a patent in a technology field $\tau$ or the export of a product $\pi$) in the activity basket of the country $c$ and the weight of that same activity with respect to the world volume, as reported in the following equation:
\begin{equation}
\label{RCA}
RCA_{ca}= \dfrac{\dfrac{X_{ca}}{\sum_{a'}X_{ca}}}{\dfrac{\sum_{c'}X_{c'a}}{\sum_{c'a'}X_{c'a'}}}.
\end{equation}
Where the element $X_{ca}$ refers to both $W_{c\tau}$ and $V_{c\pi}$, i.e. the elements of the country-green technology and country-exported product matrices (for a more detail description on how the matrices are built, we refer to the Supplementary Information [Data Features]). The next step is the computation of the binary matrices $\textbf{M} = \textbf{M}_{ca} = \{\textbf{M}_{c\tau}; \textbf{M}_{c\pi}\}$, whose elements are 1 or 0 depending on whether the value of $RCA_{ca} \ge 1$, meaning that the country $c$ is or is not competitive in activity $a$. The RCA metric is frequently used in the Economic Complexity framework to assess whether a country is a significant exporter of a product\cite{hidalgo2007product,hidalgo2009building}. The extension of its use to the patent layer\cite{pugliese2019unfolding} allows us to compare patent and export data in a coherent way in the methods we present in the following sections.

\subsection{Construction of the validated network}
\phantomsection
\subsubsection{Full technology-product network}

Starting from the binary matrices $\textbf{M}$ described above, that summarise the comparative advantage in products and technologies of different countries, a network linking green technologies to products can be derived. The method adopted here is already present in the Economic Complexity framework\cite{hidalgo2007product,pugliese2019unfolding}: the idea is to count how many countries have developed a given green technology and at the same time are competitive in the export of a product. This number is called co-occurrences\cite{teece1994understanding}. In practice, however, the co-occurrences should be suitably normalized to take into account the nested structure of the bipartite networks; the result of this process is the so-called Assist Matrix\cite{pugliese2019unfolding, zaccaria2014taxonomy}. This matrix $\textbf{A}$ is obtained from the contraction of the binary country-technology and country-product matrices. The matrix element $A_{\tau\pi}$ depends on both the year $t_{1}$ relative to the patenting of the technology $\tau$ and the $t_{2}$ of the subsequent export of product $\pi$. In formula:
\begin{equation}
\label{assist}
A_{\tau\pi}(t_{1},t_{2})= \dfrac{1}{u_\tau(t_{1})} \sum_c\dfrac{M_{c\tau}(t_{1}) M_{c\pi}(t_{2})}{d_c(t_{2})}, \text{ with } \begin{cases}
    d_{c}(t_{2}) = \sum_{\pi'}M_{c\pi'}(t_{2})\\\\
    u_{\tau}(t_{1})= \sum_{c'}M_{c'\tau}(t_{1})
\end{cases}
\end{equation} 
By counting the co-occurrences between green technologies and exported products --- while weighing them with the degree (or ubiquity) of the technology $u_{\tau}$ and the country degree (or diversification) in the exports $d_{c}$ --- each element of the matrix $A_{\tau\pi}(t_{1},t_{2})$ offers a quantitative measure on how likely is to have a comparative advantage in exporting the product $\pi$ in the year $t_{2}$, conditional on having a comparative advantage in the technology $\tau$ in the year $t_{1}$. Therefore, $t_{1}$ and $t_{2}$ indicate that it is considered the possibility that the link couples patents developed in a given year with products exported in a different year. After the computation of the Assist Matrix, we process the statistical validation of the empirical results expressed by each node $A_{\tau\pi}(t_{1},t_{2})$ through the implementation of a null model which we present in the following section.

\subsubsection{Comparison with a null model}
The matrix elements computed in Eq. (\ref{assist}) need to be validated by a statistical test able to  distinguish meaningful links from the noise and to supply a confidence level for assessing  the probability that two nodes share a statistically significant number of co-occurrences. In particular, here we rely on the filtering procedure based on the Bipartite Configuration Model (BiCM) \cite{squartini2011analytical} developed by Saracco et al.\cite{saracco2017inferring} for the projection of bipartite onto monopartite  networks, and subsequently adapted to a similar multi-partite network by Pugliese et al.\cite{pugliese2019unfolding}. It must however be noted that no absolute criteria exists for the choice of the model, and that different null models can yield different outcomes\cite{cimini2022meta}. Here, we use a null model for the binary matrices $\textbf{M}$, in which the matrices are randomised except for some constraints we impose\cite{Saracco2015} -- in this case the average degrees. The use of BiCM allows for a stricter filtering procedure with respect to other null models\cite{cimini2022meta} and takes into account the possible noise present in the input data\cite{Saracco2015,saracco2017inferring,cimini2022meta}.
This class of models is based on the maximum entropy principle\cite{Jaynes1957}, which leads to the realisation of an ensemble $\Omega$ of bipartite networks $\tilde{\textbf{M}}$, where links are random but maximize the number of possible configurations which satisfy the imposed constraints. In the present case the entropy function:
\begin{equation}
S = -\sum_{\tilde{M} \in \Omega}  P(\tilde{M})\ln P(\tilde{M})  
\end{equation}
is maximized under the constraint that the ensemble averages $\langle \dots \rangle_\Omega$ of the ubiquity of activities (i.e. of green technologies and exported products) and of countries diversification of the random  networks,  $\tilde{u}_a(t)$ and $\tilde{d}_c(t)$, must be equal the observed ones (labeled without the tilde symbol):
\begin{equation}
\begin{array}{ccc}
\langle \tilde{d}_c(t) \rangle_\Omega & = & d_c(t) \\
\langle \tilde{u}_a(t) \rangle_\Omega & = &u_a(t)
\end{array}
\end{equation}
The maximization procedure yields the probability distribution for each possible couple of nodes country-activity to be linked. Then, we use it to perform a direct sampling of the ensemble $\Omega$. The ensemble is composed of a number of realisations of the null model which necessarily depends on the threshold $p$-value with which we want to validate the links in the technology-product space. In particular, since our results are mostly set to a statistical significance of 95\%, we construct ensembles consisting of 10000 realisations of the null model. In such a way a rough but conservative estimate yields a sampling error of 5 \textperthousand. For each couple of  null model realizations $\{\tilde{M}_{c\tau}(t_{1}); \tilde{M}_{c\pi}(t_{2})\}$ related to the green technology and exported product layers, we compute the corresponding null Assist Matrix of element $\tilde{A}_{\tau\pi}(t_{1},t_{2})$ through a contraction as in equation (\ref{assist}). By doing so, we build an ensemble of 10000 realizations of the null Assist matrix. Finally, for each possible link green technology-product $\tau$-$\pi$ we compare the empirical value $A_{\tau\pi}(t_{1},t_{2})$ with the 10000 null values of that same link. We are thus able to assess the statistical significance of our results: for example, if we want to select only the links that are 95\% significant, we consider those with the experimental value higher than the corresponding null one in at least 9500 cases out of 10000. 

\subsubsection{Validation of the results for a specific time lag}
As we already stressed, the methodology at our disposal allows us to build different networks linking green technologies to exported products by varying the temporal dimension. We express the temporal dependence of the analysis through the time lag $\Delta T$ given by the difference between the year $t_{2}$ of the country-product matrix and the year $t_{1}$ of the country-green technology matrix. In particular, given the years at our disposal for the two data collections, we consider both $\Delta T= 0$ and a time lag of ten years (i.e. $\Delta T = 10$). We recall that our matrices refer to sums over 5-year intervals. For each of the two time lags considered we associate two different pairs of 5-year aggregate technology-product matrices: these are $\textbf{W}(2012)-\textbf{V}(2012)$ and $\textbf{W}(2017)-\textbf{V}(2017)$ for $\Delta T = 0$, and  $\textbf{W}(2002)-\textbf{V}(2012)$ and $\textbf{W}(2007)-\textbf{V}(2017)$ for $\Delta T = 10$, where, following equation (\ref{sumover5years}), the year is the last of the five years interval. For each couple of matrices we follow all the steps described above --- i.e. RCA, computation of the Assist Matrix, and statistical validation through the null model at a chosen $p$-value --- and we consider only those links that are statistical significant in both of them. Therefore, for instance, the links represented in Fig. \ref{Fig2} are those that with a 95\% statistical significance in both the networks  $\textbf{W}(2012)-\textbf{V}(2012)$ and $\textbf{W}(2017)-\textbf{V}(2017)$. We believe that this is an important step in order to be able to argue that the know-how of a specific technology is transmitted to a product immediately or requires a time lag of 10 years, regardless of the specific years we are considering. Moreover, it gives additional robustness to our analysis of the multi-network beyond the adoption of the null model.

\bibliography{Bibliography}

\newpage
\begin{center}
    \huge \textbf{SUPPLEMENTARY INFORMATION}
\end{center}

\section{Data Features}

\subsection{Green Patents} 

As a response to the increasing attention and concern about climate change and renewable energy generation, we are witnessing a large increase of patent applications in environment-related domains:  according to the European Patent Office (EPO), in the last years there have been around 1.5 million patent applications in sustainable technologies \cite{epo2013sustainable}. Searching for environment-related patent documents has, therefore, been a challenge, especially because in the past documents relating to sustainable technologies did not fall into one single classification. In 2013 the EPO and the United States Patent and Trademark Office (USPTO) agreed to harmonise their patent classification practices and developed the Cooperative Patent Classification (CPC) system, which encompasses five hierarchical levels spanning from 9 sections to  around 250000 subgroups and where codes starting with the letters A to H represent a traditional classification of innovative activity in technological fields, while the Y section \cite{Ysectionlink} tags cross-sectional technologies. Here in particular we employ the Y02--\textit{Technologies or applications for mitigation or adaptation against climate change} retrieved from the OECD REGPAT database \cite{maraut2008oecd}. The Y02 class consists of more than 1000 tags organised in 9 sub-classes and includes patents related to climate change adaptation and mitigation (CCMT)\footnote{According to the United Nations Environmental Program (UNEP): " Climate Change Mitigation refers to efforts to reduce or prevent emission of greenhouse gases. Mitigation can mean using new technologies and renewable energies, making older equipment more energy efficient, or changing management practices or consumer behavior''\cite{UNEP}. However, it is important to notice that mitigation does not necessarily goes hand in hand with sustainable and  "green'' practices. Some CCMTs, such as nuclear technologies, might also pose threats on the environment or be polluting.} technologies concerning a wide range of technologies related to sustainability objectives, such as energy efficiency in buildings, energy generation from renewable sources, sustainable mobility, smart grids and many others, the details of which can be found in Table 1 of the manuscript.\\
Following the notation given in the manuscript, we have matrices $\textbf{W}(t)$ from 1995 to 2019. The number of countries (i.e. the number of rows in each matrix) are 48 (see Table \hyperref[AllCountries]{S1}). The number of columns are 44 technological fields corresponding to the CPC groups listed in Table \hyperref[CPCdescription8 digits]{S2}. To build such matrices, each patent family --- i.e. each collection of patent applications covering the same or similar technical content --- counting as a unit and recorded in REGPAT is divided between all technology codes $\tau$ and all countries $c$ with which it is associated, following the procedure adopted in Napolitano et al. \cite{napolitano2020green} and Barbieri et al. \cite{barbieri2021regional}. Therefore, each element $W_{c\tau}(t)$ of the matrix represents the fraction of patent families associated with the country-technology pair $c-\tau$ in year $t$.

\subsection{Exported products} 
For the export we resort to the  UN-COMTRADE database \cite{Comtradelink}, which  provides yearly trade flows between countries in US Dollars. This information is provided at the product-level, so that it is possible to study in detail which countries are exporting a given amount of a given product in a chosen year. The products in the dataset are classified according to the Harmonized System, a hierarchical classification that allows to go from two digit  (about 100 different product chapters) up to six digits (about 5000 different product subheadings) codes. This degree of freedom is key to investigate the effect of technological innovations at different levels of detail: in fact, we move from the links that green technologies have with the export of entire product categories such as those related to the Machinery/Electrical sector to those that they have with the export of detailed single products such as electric motors. We point out that since importers' and exporters' declarations do not precisely coincide, suitable reconstruction algorithms are needed in order to achieve a coherent and cleaned dataset. In order to do so, we adopt a global Bayesian optimization approach to obtain a denoised dataset, as proposed by Mazzilli et al. \cite{mazzilli2022reconstruct}. The goodness of this procedure is empirically confirmed  by Tacchella et al. \cite{tacchella2018dynamical}, who, by employing the denoised dataset, obtained a sizeable increase in GDP forecasting  performance.\\
Finally, following the notation given in the manuscript, we have matrices $\textbf{V}(t)$ from 2007 to 2017: the number of rows, corresponding to the number of countries, is equal to 169 (see Table \hyperref[AllCountries]{S1}), while the number of columns, corresponding to the exported products, depends on the level of aggregation considered (97 in the 2-digit case, 5053 in the 6-digit one). Thus, each element $V_{c\pi}(t)$ represents the volume of exports of the product $\pi$, expressed in thousands of dollars, by the country $c$ in year $t$.

\subsection{Country list}

Depending on which step of our analysis we deal with, we consider all countries included in each collection or only those in common. In particular, the computation of the Revealed Comparative Advantage (RCA) is done separately for patents and exports, thus including all countries in the respective datasets. On the contrary, the calculation of the assist matrix is done by contracting the patent and export data over the geographical dimension, and therefore we only consider those in common. In Table \hyperref[AllCountries]{S1} we collect all the countries included in both datasets, also writing their names in different colours depending on whether they are part of the 47 common countries between the two datasets or they are only present in one of them.

\begin{table}[htbp]
\centering
\begin{tabular}{|l l l l|}
    \hline
    \multicolumn{4}{|c|}{Country full list}\\
    \hline
Afghanistan & Albania & Algeria & Andorra\\
Angola & Argentina & Armenia & \textcolor{Red}{Australia}\\
\textcolor{Red}{Austria} & Azerbaijan & Bahrain & Bangladesh\\
Belarus & \textcolor{Red}{Belgium} & Belize & Benin \\
Bhutan & Bolivia & Bosnia Herzegovina & Botswana \\
\textcolor{Red}{Brazil} & Brunei & \textcolor{Red}{Bulgaria} & Burkina Faso\\
Burundi & Cambodia & Cameroon & \textcolor{Red}{Canada}\\
Cape Verde & Central African Republic & Chad & \textcolor{Red}{Chile}\\
\textcolor{Red}{China} & Colombia & Congo & Costa Rica\\
\textcolor{Red}{Croatia} & Cuba & \textcolor{Red}{Cyprus} & \textcolor{Red}{Czech Republic}\\
Democratic Republic Congo & \textcolor{Red}{Denmark} & Dominican Republic & Ecuador\\
Egypt & El Salvador & Equatorial Guinea & Eritrea\\
\textcolor{Red}{Estonia} & Ethiopia & Fiji & \textcolor{Red}{Finland}\\
\textcolor{Red}{France} & French Polynesia & Gabon & Gambia\\
Georgia & \textcolor{Red}{Germany} & Ghana & \textcolor{Red}{Greece}\\
Greenland & Guatemala & Guinea & Guinea-Bissau\\
Guyana & Haiti & Honduras & \textcolor{Red}{Hungary} \\
\textcolor{Red}{Iceland} & \textcolor{Red}{India} & Indonesia & Iran\\
Iraq & \textcolor{Red}{Ireland} & \textcolor{Red}{Israel} & \textcolor{Red}{Italy}\\ 
Ivory Coast & Jamaica & \textcolor{Red}{Japan} & Jordan \\
Kazakhstan & Kenya & Kuwait & Kyrgyzstan \\ 
Laos & \textcolor{Red}{Latvia} & Lebanon & Lesotho\\ 
Liberia & Libya & \textcolor{ForestGreen}{Liechtenstein} & \textcolor{Red}{Lithuania}\\
\textcolor{Red}{Luxembourg} & \textcolor{Red}{Macedonia} & Madagascar & Malawi\\
Malaysia & Maldives & Mali & \textcolor{Red}{Malta}\\
Mauritania & Mauritius & \textcolor{Red}{Mexico} & Moldova\\ 
Mongolia & Montenegro & Morocco & Mozambique\\ 
Myanmar & Namibia & Nepal & \textcolor{Red}{Netherlands}\\ 
\textcolor{Red}{New Zealand} & Nicaragua & Niger & Nigeria\\
North Korea & \textcolor{Red}{Norway} & Oman & Pakistan\\
Panama & Papua New Guinea & Paraguay & Peru \\
Philippines & \textcolor{Red}{Poland} & \textcolor{Red}{Portugal} & Qatar \\
\textcolor{Red}{Romania} & \textcolor{Red}{Russia} & Rwanda & Saudi Arabia\\
Senegal & Serbia & Seychelles & Sierra Leone\\
Singapore & \textcolor{Red}{Slovakia} & \textcolor{Red}{Slovenia} & Somalia\\
\textcolor{Red}{South Africa} & \textcolor{Red}{South Korea} & South Sudan & \textcolor{Red}{Spain}\\ 
Sri Lanka & Sudan & Suriname & Swaziland \\
\textcolor{Red}{Sweden} & \textcolor{Red}{Switzerland} & Syria & Tajikistan\\
Tanzania & Thailand & Togo & Tunisia\\
\textcolor{Red}{Turkey} & Turkmenistan & Uganda & Ukraine\\
United Arab Emirates & \textcolor{Red}{United Kingdom} & Uruguay & \textcolor{Red}{USA}\\
Uzbekistan & Venezuela & Vietnam & Yemen\\
Zambia & Zimbabwe&&\\
\hline
	\end{tabular}
	\captionsetup{labelformat=empty}
	\caption{\textbf{Table S1:} All country list.\\
	Legend: "\textcolor{Red}{Red-labelled country}": included in both datasets (47 in total); "\textcolor{ForestGreen}{Green-labelled country}": included in green patents dataset only (1 in total); "Black-labelled country": included in exported products dataset only (122 in total). }
	\label{AllCountries}
\end{table}

\section{Table S2: CPC detailed descriptions}

As regarding green patents data, we have information on patent applications on 44 green technology groups. These are in turn grouped into 8 subclasses, which are reported in Table 1 of the manuscript. In Table \hyperref[CPCdescription8 digits]{S2} we report the codes and descriptions at the group aggregation level.

\begin{table}[htbp]
\centering
\resizebox{.9\textwidth}{!}{%
		\begin{tabular}{|c|c|l|}
		\hline
		\multicolumn{2}{|c|}{\textbf{CPC subclass}}&\multicolumn{1}{c|}{\textbf{Description}}\\
		\hline
		\multirow{6}*{\textbf{\textcolor{Y02A}{Y02A}}}&
		\textbf{\textcolor{Y02A}{10}}&Adaptation to climate change at coastal zones\\
		&\textbf{\textcolor{Y02A}{20}}&Water conservation\\
		&\textbf{\textcolor{Y02A}{30}}&Adapting infrastructure\\
		&\textbf{\textcolor{Y02A}{40}}&Adaptation technologies in agriculture\\
		&\textbf{\textcolor{Y02A}{50}}&in human health protection\\
		&\textbf{\textcolor{Y02A}{90}}&Indirect contribution to adaptation to climate change\\
		\hline
		\multirow{9}*{\textbf{\textcolor{Y02B}{Y02B}}}&\textbf{\textcolor{Y02B}{10}}&Integration of renewable energy sources in buildings\\
		&\textbf{\textcolor{Y02B}{20}}&Energy efficient lighting technologies\\
		&\textbf{\textcolor{Y02B}{30}}&Energy efficient heating\\
		&\textbf{\textcolor{Y02B}{40}}&Improving the efficiency of home appliances\\
		&\textbf{\textcolor{Y02B}{50}}&Energy efficient technologies in elevators\\
		&\textbf{\textcolor{Y02B}{60}}&ICT aiming at the reduction of own energy use\\
		&\textbf{\textcolor{Y02B}{70}}&Efficient end-user side electric power management\\
		&\textbf{\textcolor{Y02B}{80}}&Improving the thermal performance of buildings\\
		&\textbf{\textcolor{Y02B}{90}}&GHG emissions mitigation [Buildings]\\
        \hline
        \multirow{2}*{\textbf{\textcolor{Y02C}{Y02C}}}&\textbf{\textcolor{Y02C}{10}}&CO2 capture or storage\\
		&\textbf{\textcolor{Y02C}{20}}&Capture or disposal of greenhouse gases\\
		\hline
		\multirow{4}*{\textbf{\textcolor{Y02D}{Y02D}}}&\textbf{\textcolor{Y02D}{10}}&Energy efficient computing\\
		&\textbf{\textcolor{Y02D}{30}}&Reducing energy consumption in communication networks\\
		&\textbf{\textcolor{Y02D}{50}}&Reducing energy consumption in wire-line communication networks\\
		&\textbf{\textcolor{Y02D}{70}}&Reducing energy consumption in wireless communication networks\\
		 \hline
		 \multirow{7}*{\textbf{\textcolor{Y02E}{Y02E}}}&\textbf{\textcolor{Y02E}{10}}&Energy generation through renewable energy sources\\
		&\textbf{\textcolor{Y02E}{20}}&Combustion technologies with mitigation potential\\
		&\textbf{\textcolor{Y02E}{30}}&Energy generation of nuclear origin\\
		&\textbf{\textcolor{Y02E}{40}}&Technologies for an efficient electrical power generation\\
		&\textbf{\textcolor{Y02E}{50}}&Technologies for the production of fuel of non-fossil origin\\
		&\textbf{\textcolor{Y02E}{60}}&Enabling technologies\\
		&\textbf{\textcolor{Y02E}{70}}&Other energy conversion systems reducing GHG emissions\\
		\hline
		 \multirow{8}*{\textbf{\textcolor{Y02P}{Y02P}}}&\textbf{\textcolor{Y02P}{10}}&Metal processing\\
		&\textbf{\textcolor{Y02P}{20}}&Chemical industry\\
		&\textbf{\textcolor{Y02P}{30}}&Oil refining and petrochemical industry\\
		&\textbf{\textcolor{Y02P}{40}}&Processing of minerals\\
		&\textbf{\textcolor{Y02P}{60}}&Agriculture\\
		&\textbf{\textcolor{Y02P}{70}}&CCMT in the production process for final products\\
		&\textbf{\textcolor{Y02P}{80}}&CCMT for sector-wide applications\\
		&\textbf{\textcolor{Y02P}{90}}&GHG emissions mitigation [Production]\\
		\hline
		 \multirow{5}*{\textbf{\textcolor{Y02T}{Y02T}}}&\textbf{\textcolor{Y02T}{10}}&Road transport of goods or passengers\\
		&\textbf{\textcolor{Y02T}{30}}&Transportation of goods or passengers via railways\\
		&\textbf{\textcolor{Y02T}{50}}&Aeronautics or air transport\\
		&\textbf{\textcolor{Y02T}{70}}&Maritime or waterways transport\\
		&\textbf{\textcolor{Y02T}{90}}&GHG emissions mitigation [Transportation]\\
        \hline
        \multirow{3}*{\textbf{\textcolor{Y02W}{Y02W}}}&\textbf{\textcolor{Y02W}{10}}&Wastewater treatment\\
		&\textbf{\textcolor{Y02W}{30}}&Solid waste management\\
		&\textbf{\textcolor{Y02W}{90}}&GHG emissions mitigation [Wastewater]\\
		\hline
	\end{tabular}}
	\captionsetup{labelformat=empty}
	\caption{\textbf{Table S2:} Descriptions of environmental technology groups. In the first column (divided in turn into two sub-columns) the CPC code identifying the technology group is reported. The second column adds the corresponding group descriptions.}
	\label{CPCdescription8 digits}
\end{table}

\section{Economic Fitness \& Complexity algorithm}

In Fig. 5 of the mauscript we order the codes related to green technologies and exported products according to their level of complexity. The latter is intended as an algorithmic assessment of the number and the sophistication of the capabilities needed to be competitive in a given activity. To compute it, we use the Economic Fitness \& Complexity (EFC) algorithm  product \cite{Tacchella2012,Tacchella2013}, originally introduced for exports but also applied to green patents \cite{sbardella2018green}. More in detail, it consists of a non-linear iterative algorithm that, starting from the binary matrices $\textbf{M}_{ca}(t)$ obtained through the implementation of RCA detailed in the manuscript in the Methods section, allows to quantify the complexity of the activities $Q_{a}$ and the competitiveness of the countries, namely their fitness $F_{c}$, that perform in them. The mathematical formulation of the algorithm at each iteration $n$ is as follows:
\begin{equation}\label{EFC}
\begin{cases} 
	\tilde{F}_ {c}^{(n)} & = \sum_{a} M_{ca}Q_{a}^{(n-1)}\\ \noalign{\vskip8pt}
	\tilde{Q}_{a}^{(n)} & = \dfrac{1}{\sum_{c}M_{ca}\dfrac{1}{F_{c}^{(n-1)}}}
	
\end{cases} \;\rightarrow\; \begin{cases} 
{F}_ {c}^{(n)} & = \dfrac{\tilde{F}_ {c}^{(n)}}{	\left \langle \tilde{F}_ {c}^{(n)}  \right \rangle}_{c}\\\noalign{\vskip8pt}
{Q}_ {a}^{(n)} & = \dfrac{\tilde{Q}_ {a}^{(n)}}{	\left \langle \tilde{Q}_ {a}^{(n)}  \right \rangle}_{a}
\end{cases}
\end{equation}
where, in the left-hand bracket, the calculation of the fitness and complexity parameters for all countries and activities is shown, while in the right-hand one is the following normalisation step. The non-linear structure of the algorithm causes the activities in the baskets of less competitive countries (i.e. with low fitness) to be assigned a low level of complexity. The most competitive countries turn out to be those with more diversified activity baskets. Given the convergence properties of the algorithm, discussed in Pugliese et al. \cite{Pugliese2016conv}, we do not consider the complexity values but their rankings. In particular, the ranking are computed using the most recent 5-year aggregate matrices given the years of the data we considered in the analysis: thus, we use $\textbf{M}_{c\tau}(5, 2017)$ for green patents and $\textbf{M}_{c\pi}(5, 2017)$ for exported products.

\section{Robustness test}

In the manuscript we build the green technology-product bipartite network starting with two important preliminary steps: firstly, we summed the yearly data collections at our disposal over 5 years; secondly, depending on the time lag $\Delta T$ we consider, we select specific 5-year aggregate matrices. In particular, we select the two most recent exported product matrices available to us that do not overlap each other --- i.e. $\textbf{V}(\delta,t)=\{\textbf{V}(5,2012); \textbf{V}(5,2017)\}$, where $\delta$ corresponds to the interval of years over which the individual yearly matrices are summed up (in this case 5),  while the year $t$ explicitly indicated corresponds to the last year of the interval. Since the data collections of exported products are fixed for both time lags, we select the aggregated 5-year green patent collections depending on which of the latter we consider : therefore, we select the matrices $\textbf{W}(\delta,t)=\{\textbf{W}(5,2012); \textbf{W}(5,2017)\}$ for $\Delta T = 0$ and $\textbf{W}(\delta,t)=\{\textbf{W}(5,2002); \textbf{W}(5,2007)\}$ for $\Delta T = 10$.\\
In this section we want to show that our results do not depend on the choices of the years considered nor on the parameter $\delta$. To this aim, we conduct a robustness test in which we repeat our analysis for both different values of $\delta$ and years considered. In particular, we replicate our results for a 2-digit level of product aggregation and for the time lag $\Delta T = 0$. Considering the 10 years covered by the two 5-years summed data collections we consider in the analysis for $\Delta T = 0$ --- i.e. from 2008 to 2017 --- we create a dataset composed by 32 matrices (16 for green patents and 16 for exported products) aggregated at 3,4 and 10 years, so that $\delta = \{3, 5, 10\}$. The dataset is reported In Table \hyperref[Robtest]{S3}: each $\textbf{M}(\delta, t)$ in the table stands for a corresponding couple of technology-product matrices $\textbf{W}(\delta, t)-\textbf{V}(\delta, t)$ for which we process the full analysis, meaning RCA, assist matrix and null model computations. We consider as a benchmark of this test the 46 links validated at a 95\% level of significance in the manuscript. The results we obtain can be summarized as follows:
\begin{itemize}
    \item Considering only the aggregation over 3-year intervals, on average 73\% of the 46 links are present at a 95\% significance level. This percentage increases to 87\% if we consider a 90\% level of significance for the 3-year results.
    \item Considering only the aggregation over 4-year intervals, on average 80\% of the 46 links are present at a 95\% significance level. This percentage increases to 92\% if we consider a 90\% level of significance for the 4-year results.
    \item 85\% of the 46 links are present at a 95\% significance level for the unique pair of technology-product matrices with the 10-year time aggregation. This percentage increases to 98\% (45 links out of 46) if we consider a 90\% level of significance for the 10-year result.
\end{itemize}
Based on the above summary, we consider the robustness test successful. Therefore, we interpret the results reported in the manuscript as showing a real link of interdependence between the acquisition of green technological capabilities and the development of productive ones.

\begin{table}[hptb]
\centering
\resizebox{.98\textwidth}{!}{
\begin{tabular}{c@{}ll@{}}
\toprule
\multicolumn{1}{l}{\textbf{Time aggregation $\mathbf{\delta}$}} & \multicolumn{1}{l}{\textbf{Data collections $\mathbf{M(\delta,t)}$}} \\ \midrule
3 & \begin{tabular}[c]{@{}l@{}} M(3, 2010), M(3, 2011), M(3, 2012), M(3, 2013)\\ M(3, 2014), M(3, 2015), M(3, 2016), M(3, 2017) \end{tabular} \\
\midrule
4 & \begin{tabular}[c]{@{}l@{}} M(4, 2011), M(4, 2012), M(4, 2013), M(4, 2014)\\ M(4, 2015), M(4, 2016), M(4, 2017) \end{tabular} \\
\midrule
10 & M(10,2017)\\
 \bottomrule
\end{tabular}%
}
\captionsetup{labelformat=empty}
\caption{\textbf{Table S3:} Composition of the dataset we use for the robustness test of our results. Since we consider the time lag $\Delta T = 0$, data collections refer to both green patents and exported products.}
\label{Robtest}
\end{table}

\end{document}